\documentclass[utf8]{frontiersSCNS} 

\usepackage{url,hyperref,lineno,microtype,subcaption}
\usepackage[onehalfspacing]{setspace}
\usepackage{xspace}

\newcommand{\umgms}{mg/m$^2$\xspace}
\newcommand{\uGL}{G/L\xspace}
\newcommand{\umgL}{mg/L\xspace}

\newcommand{\ktr}{k_{tr}}
\usepackage{booktabs}
\usepackage{xcolor,colortbl}
\newcommand{\rowGrey}{\rowcolor[gray]{0.85}}

\definecolor{darkgrey}{HTML}{888888}
\definecolor{grey}{HTML}{BBBBBB}
\definecolor{lightgrey}{HTML}{F7F7F7}
\definecolor{himmel}{HTML}{E6E6E6}
\definecolor{randtextcolor}{HTML}{FFFFFF}
\definecolor{blau}{HTML}{33FFFF}

\newcommand{\mktr}{k_\text{tr}}
\newcommand{\mbase}{B}

\newcommand{\mgamma}{\gamma}

\newcommand{\mkma}{k_\text{ma}}

\newcommand{\squishlist}{
   \begin{list}{$\bullet$}
    { \setlength{\itemsep}{0pt}      \setlength{\parsep}{0pt}
      \setlength{\topsep}{0pt}       \setlength{\partopsep}{0pt}
      \setlength{\leftmargin}{1.5em} \setlength{\labelwidth}{1em}
      \setlength{\labelsep}{0.5em} } }
\newcommand{\squishend}{
    \end{list}  }

\newcommand{\bg}{\begin {eqnarray*}}
\newcommand{\eg}{\end {eqnarray*}}

\newcommand{\xOne}{x_{6mp}^{gut}}
\newcommand{\xTwo}{x_{6mp} }
\newcommand{\xThree}{x_{6tgn} }
\newcommand{\xFour}{x_{pr} }
\newcommand{\xFive}{x_{tr1} }
\newcommand{\xSix}{x_{tr2} }
\newcommand{\xSeven}{x_{tr3} }
\newcommand{\xEight}{x_{ma} }

\newcommand{\dxOne}{\dot x_{6mp}^{gut}}
\newcommand{\dxTwo}{\dot x_{6mp} }
\newcommand{\dxThree}{\dot x_{6tgn} }
\newcommand{\dxFour}{\dot x_{pr} }
\newcommand{\dxFive}{\dot x_{tr1} }
\newcommand{\dxSix}{\dot x_{tr2} }
\newcommand{\dxSeven}{\dot x_{tr3} }
\newcommand{\dxEight}{\dot x_{ma} }

\usepackage{tikz}
\usetikzlibrary{intersections}
\usepackage{pgfplots,pgfplotstable}

\usetikzlibrary{arrows,positioning,decorations.markings,decorations.pathreplacing,shapes}
\usetikzlibrary{3d,calc}
\usetikzlibrary{patterns}
\tikzstyle{vecArrow} = [thick, decoration=
	{markings,mark=at position 1 with
		{\arrow[semithick]{open triangle 60}}},double distance=1.4pt, shorten >= 5.5pt,
		preaction = {decorate},
		postaction = {draw,line width=1.4pt, white,shorten >= 4.5pt}]
\tikzstyle{doubleArrow} = [thick, decoration=
	{markings,
	mark=at position 0 with {\arrow[semithick]{open triangle 60 reversed}},
	mark=at position 1 with {\arrow[semithick]{open triangle 60}}},
   double distance=1.4pt, shorten >= 5.5pt,
   preaction = {decorate},
   postaction = {decorate}]
\tikzstyle{innerWhite} = [semithick, white,line width=1.4pt, shorten >= 4.5pt]
\pgfplotsset{
    colormap={blackwhite}{[5pt]
        rgb(0pt)=(1, 0, 0);
        rgb(1000pt)=(0, 0, 1)
    },
}
\tikzset{dot/.style={circle,fill=#1,inner sep=0,minimum size=4pt}}

\usepgfplotslibrary{fillbetween}
\pgfplotsset{compat=1.10}
\tikzset{>=latex}

\tikzset{
  markergrau/.style={
    rectangle,
    fill = himmel,
    rounded corners,
    minimum height=2.5em,
    minimum width=6em,
    inner sep=5pt,
    text centered,
  }
}

\tikzset{
  markerspecial/.style={
    rectangle,
    fill = grey,
    rounded corners,
    minimum height=2.5em,
    minimum width=6em,
    inner sep=5pt,
    text centered,
  }
}

\tikzset{
  markerellipse/.style={
    ellipse,
    fill = grey,
    minimum height=2.5em,
    minimum width=6em,
    inner sep=0pt,
    text centered,
  }
}

\tikzset{
  markernone/.style={
    draw = none,
    fill = none
  }
}

\tikzset{
  randbox/.style={
    rectangle,
    fill = darkgrey,
    minimum height=3cm,
    minimum width=2em,
    inner sep=5pt,
    text centered,
  },
}

\tikzset{ triggers/.style={ ->, thick } }
\tikzset{ triggersOut/.style={ >-|, thick } }
\tikzset{ triggersIn/.style={ |->, thick } }
\tikzset{ impacts/.style={ ->, dotted, thick } }
\tikzset{ blocks/.style={ -|, double, thick }, }

\tikzset{ arrlabel/.style={right=0.15cm} }
\tikzset{ arrlabell/.style={left=0.15cm} }
\tikzset{ arrlabelu/.style={above=0.15cm} }
\tikzset{ arrlabelb/.style={below=0.15cm} }

\newcommand{\mymarkerexp}[3]{
  \node[#3] (#2) {	
    \textbf{ \begin{tabular}{c} #1 \end{tabular} }
  }; }

\newcommand{\mymarker}[2]{ \mymarkerexp{#1}{#1}{#2} }

\newcommand{\mytrigger}[3]{ \path (#1) edge[triggers] node[arrlabel]{#2} (#3); }
\newcommand{\mytriggerIn}[3]{ \path (#1) edge[triggersIn] node[arrlabelu]{#2} (#3); }
\newcommand{\mytriggerl}[3]{ \path (#1) edge[triggers] node[arrlabell]{#2} (#3); }

\newcommand{\mytriggeru}[3]{ \path (#1) edge[triggers] node[arrlabelu]{#2} (#3); }

\def\keyFont{\fontsize{8}{11}\helveticabold }
\def\firstAuthorLast{Jost {et~al.}} 
\def\Authors{Felix Jost$^1$, Jakob Zierk$^2$, Thuy T. T.  Le$^1$, Thomas Raupach$^2$, Manfred Rauh$^2$, Meinolf Suttorp$^3$, Martin Stanulla$^4$, Markus Metzler$^2$, Sebastian Sager$^{1,5}$}
\def\Address{$^{1}$Institute of Mathematical Optimization, Department of Mathematics, Otto-von-Guericke University Magdeburg, Germany  \\
$^{2}$Department of Paediatrics and Adolescent Medicine, University Hospital Erlangen, Erlangen, Germany \\ 
$^3$Pediatric Hematology and Oncology, University Hospital ``Carl Gustav Carus'', Dresden, Germany \\
$^4$ Department of Pediatric Hemato-Oncology, Hannover Medical School, Hannover, Germany \\
$^5$ Health Campus ``Immunology, Infectiology and Inflammation (GC-$\text{I}^3$)'', Otto-von-Guericke University, Magdeburg, Germany}
\def\corrAuthor{Sebastian Sager}

\def\corrEmail{sager@ovgu.de}

\begin{document}
\onecolumn

\title[Model-based simulation of maintenance therapy of childhood ALL]{Model-based simulation of maintenance therapy of childhood acute lymphoblastic leukemia} 

\author[\firstAuthorLast ]{\Authors} 
\address{} 
\correspondance{} 

\extraAuth{}

\maketitle

\begin{abstract}%

Acute lymphoblastic leukemia is the most common malignancy in childhood. 
Successful treatment requires initial high-intensity chemotherapy, followed by low-intensity oral maintenance therapy with oral 6-mercaptopurine (6MP) and methotrexate (MTX) until 2-3 years after disease onset. 
However, intra- and interindividual variability in the pharmacokinetics (PK) and pharmacodynamics (PD) of 6MP and MTX make it challenging to balance the desired antileukemic effects with undesired excessive myelosuppression during maintenance therapy.
A model to simulate the dynamics of different cell types, especially neutrophils, would be a valuable contribution to improving treatment protocols (6MP and MTX dosing regimens) and a further step to understanding the heterogeneity in treatment efficacy and toxicity. 
We applied and modified a recently developed semi-mechanistic PK/PD model to neutrophils and analyzed their behavior using a nonlinear mixed-effects modeling approach and clinical data obtained from 116 patients. The PK model of 6MP influenced the accuracy of absolute neutrophil count (ANC) predictions, whereas the PD effect of MTX did not. 
Predictions based on ANC were more accurate than those based on white blood cell counts.
Using the new cross-validated mathematical model, simulations of different treatment protocols showed a linear dose-effect relationship and reduced ANC variability for constant dosages.
Advanced modeling allows the identification of optimized control criteria and the weighting of specific influencing factors for protocol design and individually adapted therapy to exploit the optimal effect of maintenance therapy on survival.

\tiny
 \keyFont{ \section{Keywords:} childhood acute lymphoblastic leukemia, maintenance therapy, 6-mercaptopurine, methotrexate, neutropenia,  nonlinear mixed-effects modeling, population pharmacokinetics/pharmacodynamics} 
\end{abstract}

\clearpage
\section{Introduction}

\label{intro}

Acute lymphoblastic leukemia (ALL), characterized by malignant white blood cells (WBCs) and displacement of normal hematopoiesis, is the most common childhood malignancy \cite{Hoffbrand2016}.
The treatment of childhood ALL is based on combination chemotherapy and begins with intensive, high-dose treatment for approximately 6 months (the so-called induction and consolidation therapy) followed by less-intensive, low-dose treatment (so-called maintenance therapy [MT]) that lasts for 2-3 years after disease onset.
The goal of induction and consolidation therapy is to achieve remission via lymphoblast elimination below the limit of detection, and the high intensity of these therapy elements limits further therapy intensification using conventional chemotherapy. 
Subsequent MT is essential to prevent disease relapse, and aims to maintain prolonged antileukemic activity against residual lymphoblasts, with minimal adverse events.
MT includes daily oral 6-mercaptopurine (6MP) and weekly oral methotrexate (MTX).
Blood count tests are performed regularly to ensure adequate WBC and absolute neutrophil count (ANC) suppression as a surrogate marker for antileukemic activity, without unintended excessive myelotoxicity. However, there exists no international consensus for MT dosing strategies and target levels for WBC and ANC suppression (i.e. what dose to start with, and when and how to increase or decrease chemotherapy). Empirical evaluation of different MT strategies using randomized clinical trials would be extremely challenging, due to the probably moderate effect size, the length of MT, the latency of clinically relevant endpoints, and the risk of compromising the current overall favorable outcome of childhood ALL. However, certain levels of WBC and ANC are established factors for survival, relapse or death, and other adverse events (e.g., infection), respectively. 
Therefore, a simulation model of childhood ALL MT could support the development of future MT strategies by identifying those strategies that achieve established survival factors best while avoiding established risk factors.
Mathematical models describing the pharmacokinetics (PK) of 6MP and MTX and their pharmacodynamic (PD) effects on neutrophils may help clarify the drug-exposure relationship, predict the ANC dynamics, adapt subsequent dosing amounts, and stratify patients into groups with different drug responses. 
Several PK models for 6MP \cite{Hawwa2008,Jayachandran2014,Jayachandran2015} and MTX  \cite{Panetta2002,Panetta2010,Nagulu2010,Ruhs2012,Korell2013,Hui2019,Godfrey1998} have been published, but not all have been developed with low-dosage treatments and validated in the pediatric population.
To the best of our knowledge, there are only three publications \cite{Jayachandran2014,Le2018a,Karppinen2019} in which some of the PK models or their simplifications were linked to transient PD compartment models \cite{Upton2014}.
The models were individually fitted to WBC counts and different prediction and optimization studies were conducted.

Here, we developed a population PK/PD model for 6MP/MTX maintenance treatment of ALL in children based on the approach used by \cite{Le2018a} with a modified underlying PK model. 
As ANCs are the best established risk and survival factors, we adapted the model to predict ANCs instead of WBCs.
The model was fitted to and validated on a dataset consisting of weekly ANC measurements obtained from 116 patients treated with daily oral 6MP and weekly oral MTX over an average of 459 (range, 200-581) days.  
We started our investigations with a PK/PD model considering 6MP and MTX but the constant administration ratio hampered the identification of seperate PD effects.
Further, the PK of MTX had no significant impact on the improvement of the model fitting, similar to the findings in \cite{Karppinen2019}.
Thus, the final model only contains the PK of 6MP.
We take this issue up in the discussion.
Then, for each patient, we simulated different therapy protocols (6MP and MTX dosing regimens), and compared the resulting predictions.

\clearpage
\section{Patients and methods}


\subsection{Data}
\label{subsec:1}

The data used in this study were obtained retrospectively from 116 children who were diagnosed with de novo ALL at university hospitals in Erlangen and Dresden and treated according to the AIEOP-BFM 2000 and 2009 protocols. A subset of this data set (WBC counts from 9 patients) was used and described similarly in a previous study \cite{Le2018a}.
Patients were eligible if they were diagnosed with precursor B-cell or T-cell ALL, negative for the BCR-ABL- and MLL-AF4 translocations, and started MT (i.e., did not experience relapse before the end of consolidation therapy and did not undergo stem cell transplantation). 
During MT administered according to the AIEOP-BFM 2000 and 2009 protocols, patients received oral chemotherapy with daily 6MP and once-weekly MTX until 2 years after ALL diagnosis. 
During MT, chemotherapy was applied to achieve antileukemic activity against lymphoblasts below the limit of detection. As a surrogate for antileukemic activity, WBC and ANC were measured regularly, with ANC $<$2~\uGL, being correlated to a significantly better relapse-free survival \cite{Schmiegelow2014}, and ANC $<$0.5~\uGL being an indicator of excessive myelosuppression. 
The target range for the WBC count was 1.5--3~\uGL.
The chemotherapeutic dose was reduced when cell counts fell below the lower limits (WBC count $<$1.5~\uGL, ANC $<$0.5~\uGL, lymphocyte count $<$0.3~\uGL, and platelet count $<$0.05~\uGL) or liver toxicity was suspected. 
For each patient included in the analysis, data regarding the following variables were recorded: gender, age, weight, height, body surface area (BSA), prescribed 6MP and MTX dosages (absolute and per BSA), WBC count, platelet count, lymphocyte and neutrophil counts, and therapy interruptions.
In this study, we focused on 5897 ANCs and 6640 WBC counts, disregarding measurements of other cell types.
We used both WBC counts and ANC separately and compared the accuracy of the resulting mathematical models.
In all, 1150 ANC and 1289 WBC count measurements were excluded due to concurrent high C-reactive protein (CRP) levels indicating periods in which patients probably suffered from an infection. 
More precisely, we excluded measurements in the interval from two weeks before until two weeks after CRP levels of $>$5~\umgL were recorded.
Among the remaining 4747 ANC measurements 56\% were below the ANC threshold of 2~\uGL, only 2\% were below 0.5~\uGL, and 54\% were in the ANC target range 0.5--2~\uGL.
The demographic and clinical characteristics of the pediatric ALL population are shown in Table \ref{tab:datasetMedian}.

\subsection{Nonlinear mixed-effects modeling and parameter estimation}
\label{subsec:NLMEmodeling}

The nonlinear mixed-effects (NLME) modeling \cite{Bonate2011} was based on the PK/PD model of \cite{Le2018a}.
It describes the absorption of both drugs through the gastrointestinal tract into the plasma after oral administration and their metabolization to their active forms.
The MTX metabolites MTXPG$_2$ to MTXGP$_7$ inhibit several enzymes responsible for DNA synthesis \cite{Panetta2002}.
The active form of 6MP, 6-thioguanine nucleotides (6-TGNs), is incorporated into the DNA \cite{Hawwa2008}.
Thus, both drugs negatively affect the hematopoiesis of neutrophils, which is described by a chain of five compartments.
The first compartment represents the proliferating stem cells. It is negatively affected by 6MP and MTX via a linear PD term with one joint PD parameter.
The next three transit compartments describe the maturation process until mature neutrophils are released into the circulating blood (last compartment).
Further details about this model have been described \cite{Le2018a}. 
Modifications of the model and the underlying mathematical equations will follow in Section~\ref{sec:results}.

During the model development, we replaced the 6MP PK model of \cite{Jayachandran2014} with the PK model described by \cite{Hawwa2008} to obtain a better response to 6MP dosage. 
The model contained the BSA as a covariate in the clearance and thus provides individualized PK profiles. The PK model of \cite{Jayachandran2014} was validated on concentration data of 8 patients (adults) from \cite{Hindorf2006}.
However, the simulated 6-TGN concentrations coincided with data from pediatric patients reported by \cite{Hawwa2008}; hence, it was a priori unclear which would give better results.
We also tested the influence of weekly MTX administration by either ignoring or considering the administrations and their resulting concentrations through the MTX PK model with a second PD parameter during model fitting.
We also tested the myelosuppression model from \cite{Jayachandran2014}, which contained a different feedback term for ANC recovery, but the accuracy decreased and this line of research was not further investigated. 

The steady state of neutrophils $\text{Base}$, the transition rate $\mktr$, the feedback term $\mgamma$, and the PD effect slope were defined as parameters.
Interindividual variability (IIV) was assumed as log-normally distributed for all four parameters, and the residual variability was estimated using a proportional error model.
A linear residual variability model was tested, but resulted in a reduced model accuracy.

\subsection{Out-of-sample validation}
\label{subsec:3}


The reliability of the final population PK/PD model was tested via out-of-sample cross-validation.
For each patient, the first 70\% of ANC measurements were used for parameter estimation and the final 30\% were used to evaluate the model predictions.
Model accuracy and predictability were evaluated using the root mean squared error (RMSE) and the mean absolute error (MAE).

\subsection{Simulation study}
We compared individual simulated minimal, median, and maximal ANCs resulting from the application of different dosing regimens (MT dosage over time). 
The choice of the different doses described in Table~\ref{tab:Protocols} was based on ALL treatment protocols (AIEOP-BFM 2009 with EudraCT number 2007-004270-43, NOPHO-ALL 2008-003235-20, and UKALL 2010-020924-22). 
In particular, we sought to investigate the relationship between an increased total amount of chemotherapy (higher dosage) and plausibly reduced ANC in the \textit{in silico} simulations.
Throughout, we used the fitted models (estimated model parameters) from Section~\ref{subsec:NLMEmodeling} and only varied the chemotherapy dosage.
The simulated ANC values were obtained from the individual actual measurement time points.

\subsection{Software}

The population PK/PD analysis was performed with the nonlinear mixed-effects modeling program NONMEM 7.4 (ICON Plc., Dublin, Irland) \cite{Beal2009}. 
The parameters were estimated using the first order conditional estimation method with interaction.
Standard errors were computed with the \$COVARIANCE step in NONMEM.
Pirana (Certara, Princeton, USA) was used for the generation of the visual predicitive check with \textit{auto\_bin} option.
The simulations in section \textit{out-of-sample validation} and \textit{simulation study} were performed with the ODE integrator CVodes (Sundials, Lawrence Livermore National Laboratory, Livermore) \cite{Hindmarsh2005} interfaced to CasADi (Optimization in Engineering Center [OPTEC], K.U. Leuven) \cite{Andersson2019}.

\clearpage
\section{Results}
\label{sec:results}

\subsection{Mathematical model}

Table~\ref{tab:ModelComparison} shows RMSEs, MAEs, and final objective function values for four different parameter estimations.
Here, we compared the usage of different PK/PD models and estimation based on either WBC counts or ANCs. 
First, the explicit consideration of MTX only had a minimal/non-significant effect on the model accuracy, so we fixed it to the ratio 2.5:1 between 6MP and MTX and neglected the PK of MTX in the following.
Second, our results showed that the use of the PK model of \cite{Hawwa2008} increased the sensitivity of the PD effect and the model accuracy compared to the 6MP PK model of \cite{Jayachandran2014}. 
Third, ANC measurements resulted in higher accuracy than did WBC measurements.

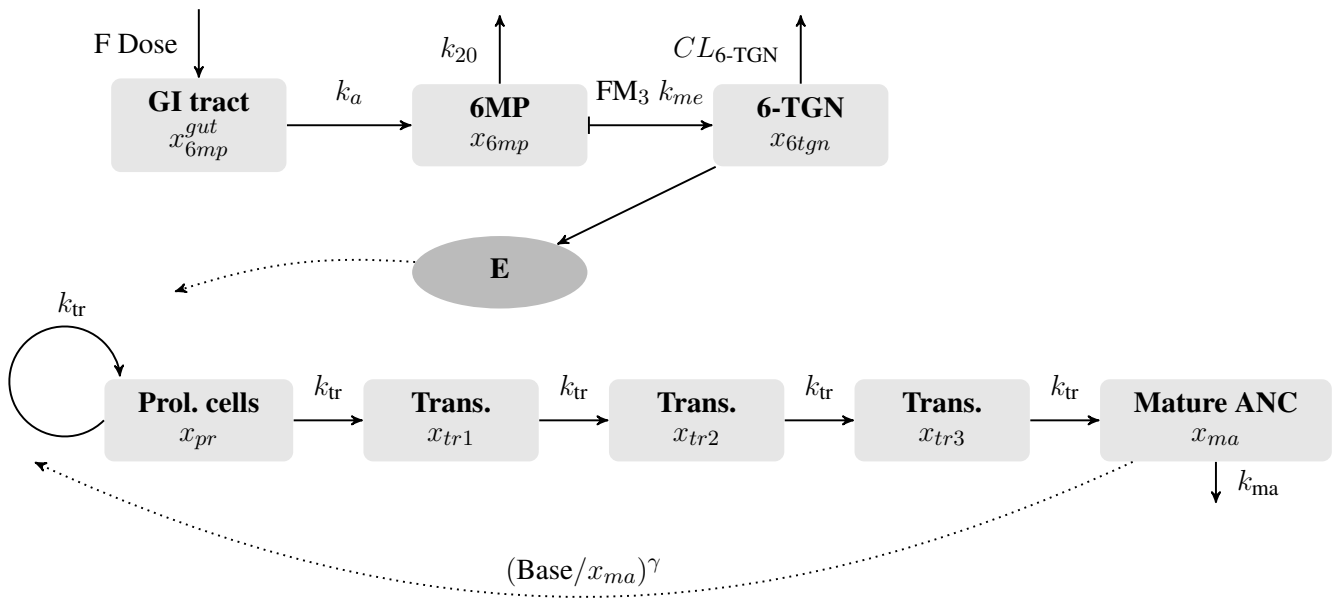
\begin{figure}[h!]
\begin{center}

\resizebox{1.\textwidth}{!}{
\begin{tikzpicture}[->,>=stealth']
  \mymarkerexp{Prol. cells \\ $\xFour$}{xpr}{markergrau}
  \mymarkerexp{Trans.  \\ $\xFive$}{xtr}{markergrau, right=1cm of xpr}
  \mymarkerexp{Trans.  \\ $\xSix$}{xtr2}{markergrau, right=1cm of xtr}
  \mymarkerexp{Trans.  \\ $\xSeven$}{xtr3}{markergrau, right=1cm of xtr2}
  \mymarkerexp{Mature ANC \\ $\xEight$}{xma}{markergrau, right=1cm of xtr3}
  \mymarkerexp{}{death}{markernone, below=0.6cm of xma}
  \mymarkerexp{}{feedback}{markernone, left=0.8cm of xpr, yshift=-1.3cm}
  \mymarkerexp{}{feedbackbeta1}{markernone, left=1.cm of xpr, yshift=-.5cm}
  \mymarkerexp{}{feedbackbeta2}{markernone, left=-1.cm of xtr, yshift=-1.4cm}
  \mymarkerexp{}{feedback2}{markernone,  above=.4cm of xpr,yshift=.6cm,xshift=-.6cm}

  \mytriggeru{xpr}{$\mktr$}{xtr}
  \mytriggeru{xtr}{$\mktr$}{xtr2}
  \mytriggeru{xtr2}{$\mktr$}{xtr3}
  \mytriggeru{xtr3}{$\mktr$}{xma}
  \mytrigger{xma}{$\mkma$}{death}
 
  \draw [triggers] (xpr.west) arc (-45:-355:8mm) node[pos=0.75,above] {$\mktr$};

  \path[impacts] (xma) edge[bend left=25] node[above] {$(\text{Base}/x_{ma})^\gamma$}  (feedbackbeta1);


  \mymarkerexp{GI tract \\ $\xOne$}{x1}{markergrau, above=3cm of xpr}
  \mymarkerexp{6MP \\ $\xTwo$}{x2}{markergrau, right=1.8cm of x1}
  \mymarkerexp{6-TGN \\ $\xThree$}{x3}{markergrau, right=1.8cm of x2}
  \mymarkerexp{}{dummyx2}{markernone, below=.75cm of x3}
  \mymarker{E}{markerellipse, below=1.cm of x2}
  \mymarkerexp{}{input}{markernone, above=1.cm of x1}
  \mymarkerexp{}{death2}{markernone, above=1.cm of x2}
  \mymarkerexp{}{death3}{markernone, above=1.cm of x3}
  \mymarkerexp{}{dummyx3}{markernone, , below=1.5cm of x2, xshift=-1.cm}

  \mytriggerl{input}{F Dose}{x1}
  \mytriggeru{x1}{$k_a$}{x2}


  \mytriggerl{x2}{$k_{20}$}{death2}
  \mytriggerl{x3}{$CL_{\text{6-TGN}}$}{death3}
 
  \mytriggerIn{x2}{$\text{FM}_3 \ k_{me}$}{x3}

  \mytriggerl{x3}{}{E}
  \path[impacts] (E) edge[bend right=11] (feedback2);

\end{tikzpicture}
}
\end{center}
\caption{Visualization of the final compartment model used for the population PK/PD analysis. 
The PK model was published by \cite{Hawwa2008} and the myelosuppression model by \cite{Le2018a}.}
\label{fig:Schematic}    
\end{figure}

As a result, we identified one PK/PD model which described the clinical data best. This model was formulated as a system of ordinary differential equations:

\begin{equation}\label{eq:FinalPKPDmodel}
\begin{aligned}
{\dxOne}(t) &=  -k_a \ \xOne(t) + F \ u(t)   ,\\
{\dxTwo}(t) &=  \phantom{-} k_a \ \xOne(t) - k_{20} \ \xTwo(t),\\
{\dxThree}(t) &=\text{FM}_3 \ k_{me} \ \xTwo(t) - \text{CL}_{6tgn}(\text{BSA}) \ \xThree(t)\\
{\dxFour}(t)& = k_{prol} \ \xFour(t)	 \ (1-E_{drug}) \ \left(\frac{\text{Base}}{x_{ma}(t)}\right)^{\gamma}-k_{tr} \ \xFour(t),\\
{\dxFive}(t)& = k_{tr} \ (\xFour(t)- \xFive(t)),\\
{\dxSix}(t) &= k_{tr} \ (\xFive(t)- \xSix(t)),\\
{\dxSeven}(t)& = k_{tr} \ (\xSix(t)- \xSeven(t)),\\
{\dxEight}(t)& = k_{tr} \ \xSeven(t)- k_{ma} \ \xEight(t)
\end{aligned}
\end{equation}
with the BSA-dependent clearance
\begin{align}
\text{CL}_{6tgn}(\text{BSA}) & = 0.00914 \ (\text{BSA})^{1.16},
\end{align}
the linear pharmacodynamic effect
\begin{align}
E_{drug}  & = \text{slope} \ x_{6tgn},
\end{align}
and the patient-specific amount $u(t)$ of 6MP (implemented as point administration in NONMEM). 
The PK of 6MP is described by a three compartment model altered from \cite{Hawwa2008}.
A fraction of the orally administered 6MP dosage enters the GI tract where bioavailable 6MP is absorbed to the central compartment with the first order rate $k_a$.
In the central compartment 6MP is eliminated by $k_{20}$. 
The elimination also comprises metabolization of 6MP by the rate $k_{me}$ out of which a fraction FM$_3$ is metabolized to the active form 6-TGN.
6-TGN is then cleared by the BSA-dependent clearance term $\text{CL}_{6tgn}$.
The hematopoiesis of neutrophils is described by a chain of five compartments with equivalent transition rates $k_{tr}$ representing the mean maturation time of the neutrophils. 
The proliferation rate of hematopoietic stem cells is equivalent to the transition rate $k_{tr}$ guaranteeing homeostasis \cite{DeSouza2018}.
Deviations from the neutrophil baseline Base are compensated by the feedback regulation $(\mbase/x_{ma})^\gamma$.
Matured neutrophils die by the process of apoptosis with the rate $k_{ma}$.
A schematic representation of the model is shown in Figure~\ref{fig:Schematic} and model constants are listed in Table~\ref{tab:Constants}.
As no PK biomarkers were measured in the examined dataset, we relied on published PK models and individualized the PD model with respect to individual sets of PD parameters.
The compartment model of \cite{Hawwa2008} had a comparable representation of the absorption and metabolic pathway of 6MP compared to the model of \cite{Jayachandran2014}.
However, the metabolic transformations are described by first order kinetics instead of Michaelis--Menten kinetics and the elimination is described by a BSA-dependent clearance term.

\subsection{Parameter estimation}

\begin{figure}[h!]
\begin{center}
\includegraphics[width=1.\textwidth]{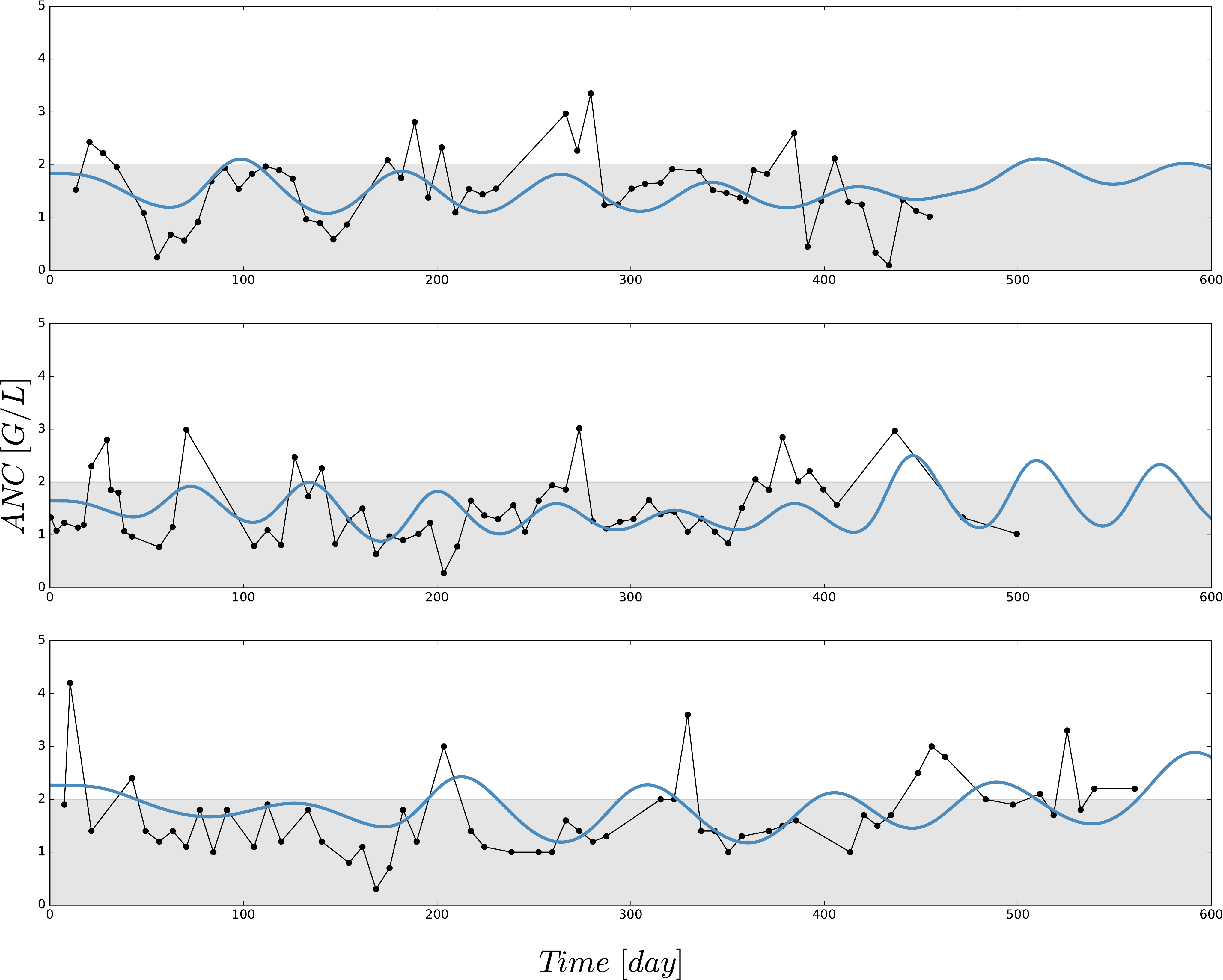} 
\end{center}
\caption{Exemplary comparisons of observed (black) absolute neutrophil counts (ANCs) and individual simulation results (blue), based on the newly proposed mathematical model and a nonlinear mixed-effects parameter estimation.
In addition to a visual match between values and quite well captured trends (compared with the indicators in Table~\ref{tab:ModelComparison}) one can clearly see oscillations of ANCs in both the observed and simulated data.
}
\label{fig:ExemplaryFittedModels}
\end{figure}

Figure~\ref{fig:ExemplaryFittedModels} shows exemplary comparisons of observed clinical and simulated ANCs. 
The simulated ANC trajectories had an oscillating nature. 
They represented the clinical ANCs quite well in the average and capture trends toward larger or smaller ANC values. 
However, they did not oscillate as strongly as the measured values.
Several reasons were responsible for the observed ANC oscillations such as aberrant hematopoiesis, chemotherapeutic dose adaptations, infections or measurement errors. 
This exemplary behavior was representative of the entire data set of 116~patients. 
The visual predictive check plot in Figure~\ref{fig:VPC} shows the good agreement of model response and measurements for the median (solid line) and 97.5th percentile (dashed line) with a slight underprediction of the model for low ANC values.
The 95\% confidence interval of the model simulation median was very thin, indicative of high prediction accuracy.
The fixed effect estimate for the ANC steady state was slightly higher than the target range limit of 2~\uGL.
The estimated transition rate of 0.148 resulted in a mean maturation time (MMT = $n_{tr}$/$k_{tr}$) of 487 hours (20.3 days) \cite{DeSouza2018}.
The interindividual variability and residual error were within reasonable ranges.

\begin{figure*}[h!]
  \includegraphics[width=1.\textwidth]{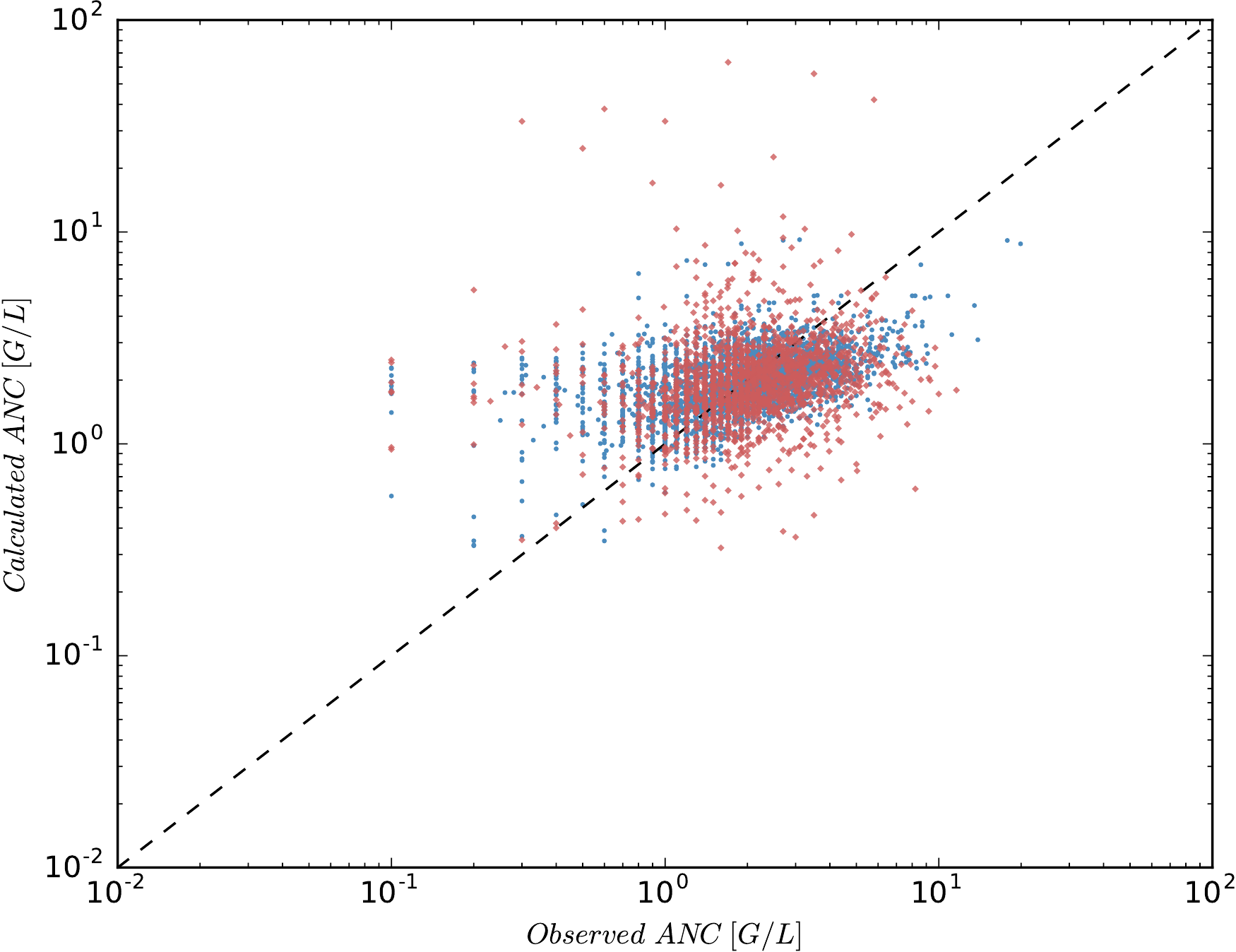}
\caption{Goodness-of-fit plot depicting observed versus individually calculated absolute neutrophil counts (ANCs) for 116 patients. 
Blue markers show in-sample ANCs (first 70\% of observed ANCs) used for parameter estimation.
Models were cross-validated using 30\% out-of-sample observed ANCs (red).
}
\label{fig:GoF}    
\end{figure*}

The goodness-of-fit plot in Figure~\ref{fig:GoF} shows the results of out-of-sample cross-validation. It reflected reasonable model accuracy for fitted (blue) and predicted (red) ANC measurements with spreading around the line of identity because the model was not able (and not intended) to hit the lower and upper peaks of the measurements. 
The values of estimated model parameters both for the in-sample and out-of-sample calculations are shown in Table~\ref{tab:FinalParameterEstimates}. 
The values of all four parameters coincided well, with a slightly reduced slope and increased $k_{tr}$ value for the estimates based on 70\% of the ANC. 
The interindividual variability for the slope was significantly larger.
To evaluate the model accuracy, we calculated the median and standard deviation of the individual MAEs and RMSEs, showing the expected decrease in accuracy for out-of-sample predictions.


\subsection{Simulation}

Figure~\ref{fig:Boxplots} shows boxplot results for an \textit{in silico} simulation study based on the 6 different treatment protocols (including the real clinical data) from Table~\ref{tab:Protocols}.
We want to stress three main observations.

First, a comparison of the first two entries of the three boxplots confirmed an already known result. 
The personalized models could reproduce the clinical ANC data on average quite well, with the exception of extreme values.
Given the similarity of simulated and observed median values, we continued with an objective comparison only of the simulated results (protocols 2--6).

Second, a comparison of the protocols 3--6 (25, 50, 75, and 100~\umgms BSA 6MP) showed a significant and linear dosage-effect relationship with respect to the total amount of 6MP administered, which is, of course, proportional to the daily dose.
The minimal and median ANC values decreased linearly, when daily dosing was increased linearly. However, the maximal ANC values were in the same range for all four protocols.


Third, a comparison of protocol~2 (the simulation of the real treatment) and protocols~3 and 4 (which gave lower and upper bounds on the total amount of administered 6MP in protocol~2, respectively) showed that the median ANC value of protocol~2 was indeed bounded by the two other values, however, for significantly lower minimal and higher maximal ANC values.
Figure~\ref{fig:exampleTraj} shows an exemplary comparison of protocols~2--6 for one patient, highlighting differences in the induced oscillations of ANC. Similar plots for all 116 patients are provided in the online supplement.


\section{Discussion}

\subsection{Mathematical model}

We developed and fitted a population PK/PD model to assess the ANC dynamics during 6MP/MTX treatment, get a better understanding of dose adjustments, and identify solutions to the challenges that arise during MT.
During the model development process we also fitted the model to WBC measurements.
The resulting MAEs and RMSEs were worse compared to the values resulting from ANC measurements.
This is probably due to the fact that WBCs comprise different cell lineages, with additional physiological effects that are not accounted for in the mathematical model. 
In future studies, the current model might be extended to further cell lineages.
The models brought forth by \cite{Quartino2012,Fornari2019} might serve as a basis and drive the modeling process from a semi-mechanistic approach toward a more mechanistic one.

In addition to using a population estimation approach and applying it to ANC instead of WBC, two modifications brought forth by \cite{Le2018a} were shown to yield better results. 
First, the 6MP PK model of \cite{Jayachandran2014} was replaced by that of \cite{Hawwa2008}.
The first order kinetics in the PK model of \cite{Hawwa2008} compared to the Michaelis--Menten terms in the PK model of \cite{Jayachandran2014} resulted in more significant concentation changes with altered drug amounts consequently in a more sensitive PD effect.
Second, the MTX PK model was completely omitted as the constant ratio of administered 6MP and MTX prevents a differentiation of seperate PD effects. 
Further studies with measurements of drug concentrations, metabolites and clinical effects as cell counts would push forward the development of a mathematical model additionally including the PK of MTX to provide two distinct PD effects and to account for varying ratios of 6MP to MTX. 
For the currently available data, our new model, which indirectly agglomerates the effects of 6MP and MTX, appears to be a good choice (compare for Table \ref{tab:ModelComparison}).

; second, the MTX PK model was completely omitted. 
Further studies with measurements of drug concentrations and metabolites would help develop a mathematical model that would account for varying ratios of 6MP to MTX. 
For the currently available data, our new model, which indirectly agglomerates the effects of 6MP and MTX, appears to be a good choice (\ref{tab:ModelComparison}).


\subsection{Model parameter estimates}

Looking at the resulting model parameter estimates listed in Table~\ref{tab:FinalParameterEstimates}, the question arises as to how these values relate to known biological properties of hematopoiesis and myelosuppression and to other values from the literature.
The estimated ANC steady state value Base was below the normal ANC range for children, but still higher than the desired ANC range of 0.5--2~\uGL. 
Without treatment, the model-based ANCs would increase to normal patient-specific steady states. 
Thus, low ANC values were induced via MT or some of the aforementioned external events. 

The estimated fixed-effects parameter value of the transition rate $\mktr=0.148$ was comparable with the published mean value ($\bar \mktr =0.1431$) obtained from eight pediatric ALL patients from Riley Hospital for Children in Indianapolis \cite{Jayachandran2014}.
For better interpretability, the transition rate parameter $\mktr$ can be transformed to the MMT ($n_{tr}$/$k_{tr}$) of the neutrophils.
The estimated MMT in our study, as well as the MMT from Jayachandran \textit{et al.}, are extremely high and do not coincide with biological findings of 3.9 days obtained by \cite{Hearn1998}.
This mismatch is a large disadvantage of the model as it fails to comply with biological properties, leading to falsely characterized physiological mechanisms and thus reduced model reliability. 
Jayachandran \textit{et al.} did not discuss this issue, but a similar observation was made by Craig and colleagues (2016) who determined an estimated proliferation time of 26 days (\cite{Craig2016}).
In their work, the authors further presented model modifications to obtain a more realistic maturation time of 3.9 days.
For this value we performed two parameter estimations with either Base as a parameter or fixed to 5 resulting in promising dynamics but worse RMSEs and MAEs.
In future studies, the falsely determined MMT and possible model limitations for continuous low-dose treatments should be further investigated. 

The feedback parameter ($\mgamma$) is significantly higher compared with published values \cite{Friberg2002}, indicating a stronger feedback mechanism during the daily chemotherapy over a long period.
This is the first time estimated slope values of the linear PD function from the PK model of \cite{Hawwa2008} are presented; thus there are no available comparisons.

\subsection{Simulation results}

\begin{figure}[h!]
\begin{center}
\includegraphics[width=.3\textwidth]{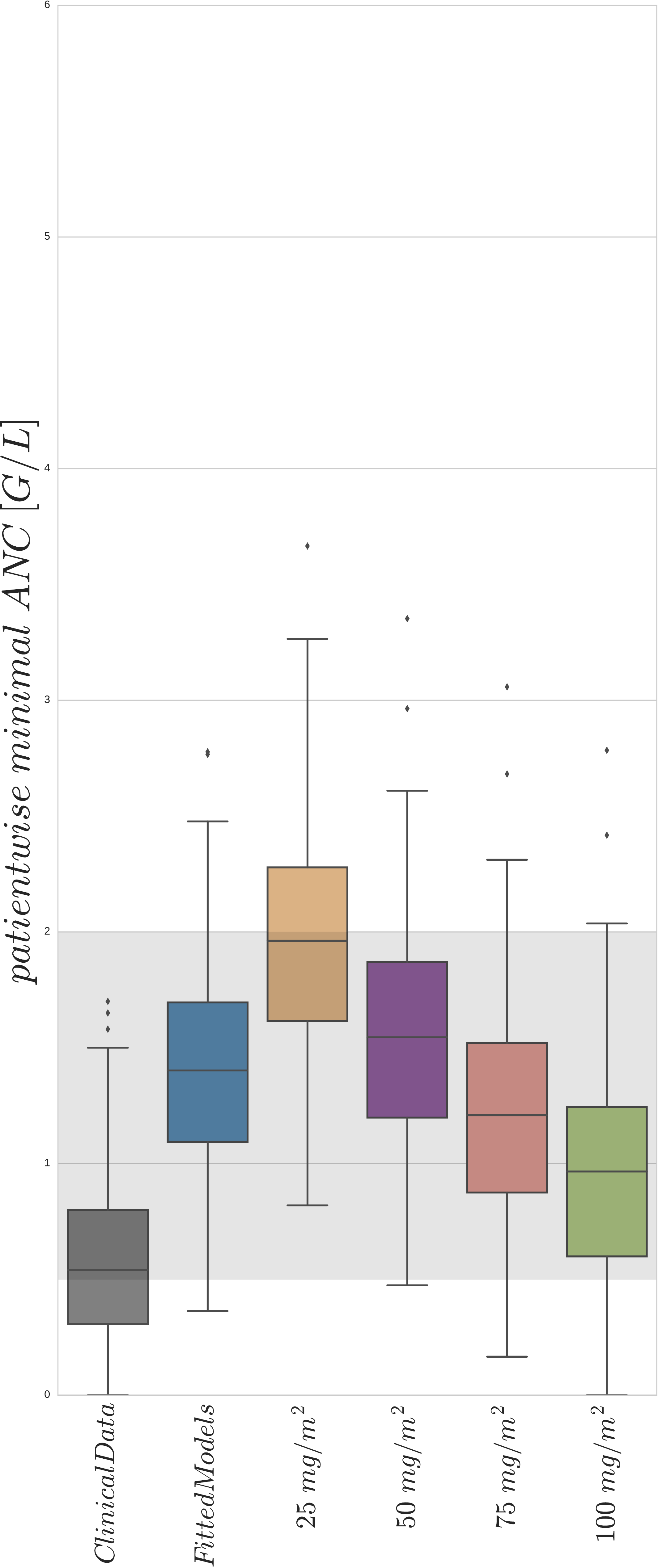} 
\includegraphics[width=.3\textwidth]{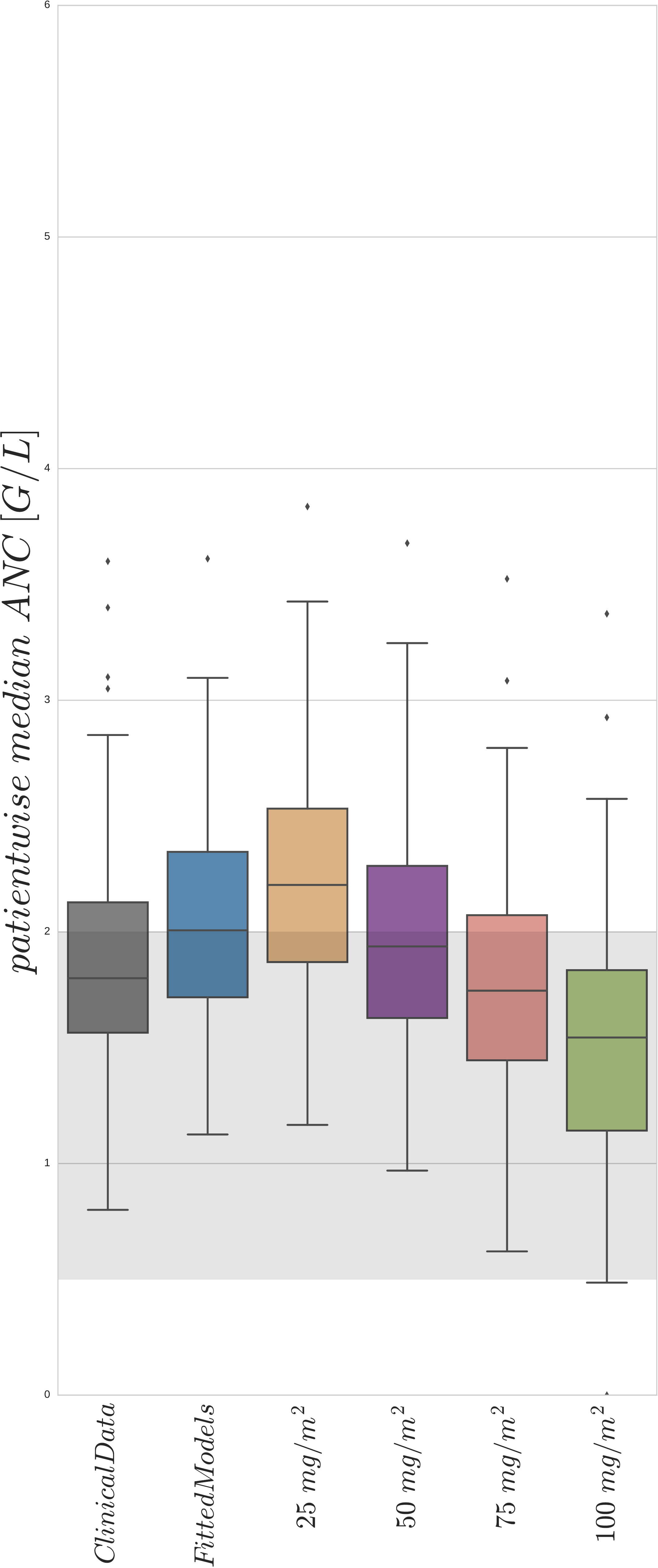}
\includegraphics[width=.3\textwidth]{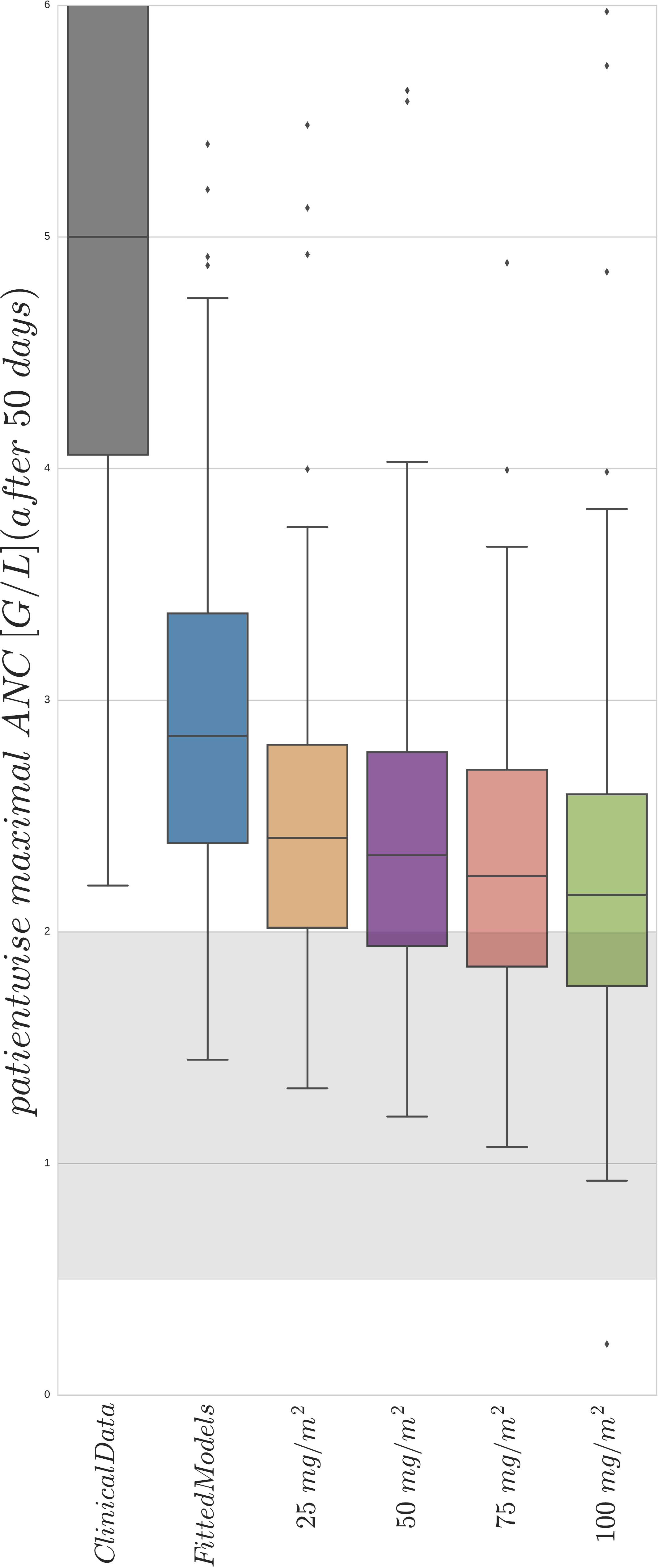} 
\end{center}
\caption{Boxplots of minimal, median, and maximal (from left to right) individual ANCs for all 116 patients. 
Shown are values for the 6 different protocols from Table~\ref{tab:Protocols}, observed for the first column and simulated for protocols 2--6. 
The target range (0.5-2.0~\uGL) of the NOPHO/UK treatment protocol is shown as the gray background. 
Horizontal lines within the boxes are the medians, the upper and lower box limits are the first and third quartiles of the data, respectively. 
The whiskers indicate an even larger confidence region of these quartiles plus/minus 1.5-times the interquartile range.
Beyond the whiskers, data are considered as outliers and are plotted as individual points. 
For the columns representing 25~\umgms to 100~\umgms, the total amount of 6MP administered is increasing. 
The median average individual daily doses actually administered for protocols~1 and 2 were $43.15\pm10.5$~\umgms.
}
\label{fig:Boxplots}
\end{figure}

The newly developed mathematical model enables us to perform a virtual comparison of different treatment protocols. 
The boxplots in Figure~\ref{fig:Boxplots} show several interesting results.

First, the median and standard deviation of actual ANC measurements were very accurately matched by the simulation using the estimated parameters (compare first two entries in middle boxplot of Figure~\ref{fig:Boxplots}). 
The variability in the clinical data, particularly toward larger ANC values, was larger, as expected and as shown in a comparison of the minimal and maximal boxplots. 
This variability is biologically and clinically very plausible due to the aforementioned external events and uncertainties, although periods of severe infections were already excluded. 
The reproducibility of the median and avoidance of over-fitting of the extreme values are in our opinion good properties of a mathematical model.
Given this good correspondence between cross-validated data and simulations, we felt encouraged to compare simulations of different treatment protocols as specified in Table~\ref{tab:Protocols}. 
Note, however, that generalizations of mathematical models personalized for data from one protocol to another have to be considered with extreme care (compare the discussion for acute myeloid leukemia models by \cite{Jost2019}). 
The impact of model variations on the outcome of simulation studies is usually significant. 
We tested the value of fixing the $\mktr$ parameter to represent a biologically plausible MMT of 3.9 days. This decreased the model accuracy (which is why the results are not included here), but still led qualitatively to the same subsequent effects.

\begin{figure*}[h!]
\begin{center}
  \includegraphics[width=1.\textwidth]{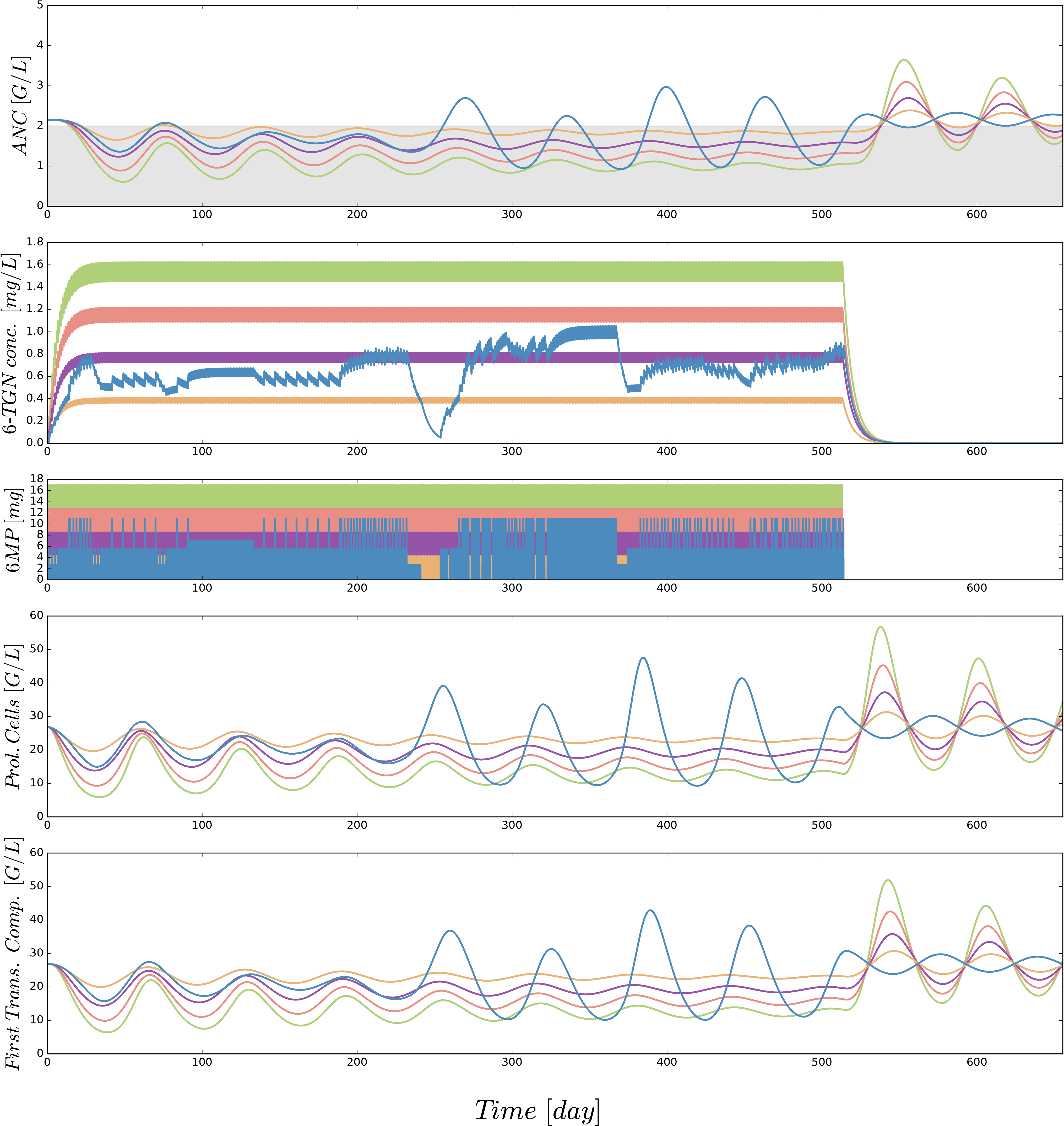}
\end{center}
\caption{Simulated trajectories for 5 different protocols from Table~\ref{tab:Protocols} and an exemplary patient. 
Colors of the trajectories are identical to those used in Figure~\ref{fig:Boxplots}.
The linear increase in dosing from 25~\umgms to 100~\umgms forces the neutrophils (ANC) to lower peak values and a smaller drug-induced steady state value at the end of treatment.
The actual dosage administered to the patient (blue) ranged between the 25~\umgms and 50~\umgms protocols and resulted in similar ANC dynamics.
At approximately day 240, the actual dosing was stopped for a short period, inducing stronger ANC oscillations in the subsequent treatment period.
This observation is even stronger regarding the proliferating cells as well as cells in the first transit compartment.
Interestingly, these oscillations also continued for some time after the end of treatment. 
}
\label{fig:exampleTraj}    
\end{figure*}

Second, an approximately linear decrease in minimum and median values could be observed as the dosage increased linearly from 25~\umgms to 100~\umgms. 
Again, this linear dose-effect relationship seems biologically plausible. 
The maximum ANC values were approximately identical, though. 
For some individual examples such as those shown in Figure~\ref{fig:exampleTraj}, the maximal ANC value also decreased. 
However, other (see online supplement) stronger oscillations led to identical or even higher maximal ANC values. 
This effect is due to a feedback mechanism that may lead to increased proliferation for reduced ANC which leads to larger ANC values after some delay.


Third, a tendency for higher oscillations for treatments with pauses and changes in dosage was seen in a comparison of the simulated actual treatment protocol~2 and the constant administrations of protocols 3 and 4, which used lower/higher total amounts of 6MP. 
Again, an example of this can be seen in Figure~\ref{fig:exampleTraj}.
For many patients, the simulations also showed an oscillating behavior also after the end of MT. 
Often amplitudes were higher than during MT when damping via 6MP administration took place, and only slowly oscillated into a steady state (see online supplement for examples). 
Such oscillations could also be seen in the clinical data (compare Figure~\ref{fig:ExemplaryFittedModels}). 
Interestingly, less damping occurred for protocols with higher doses and pauses, leading to larger oscillations that continued over time and made decision making more difficult.
The connection between \textit{model-intrinsic} and \textit{chemotherapy-induced} oscillations should be assessed in detail in future studies.
A stability analysis \cite{Edelstein2005} of the steady state could be performed (e.g., similar to \cite{Stiehl2014} and \cite{Tetschke2018}) to assess the theoretical properties of the model and relate them to the physiological behavior of neutrophils. 

\section{Conclusion}

We presented a novel nonlinear mixed-effects model that combined PK/PD and myelosuppression for ALL MT among children who received 6MP and MTX and was cross-validated on a data set of 4747 ANC measurements obtained from 116~patients. 
A comparison with alternative modeling approaches and using WBC counts instead of ANCs showed the benefit of this model. 
We could show a linear dose-effect relationship superimposed with fluctuations of varying magnitude. Mathematical simulations will allow to improve the understanding of intrinsic and extrinsic influence factors on oscillations.
In the future and based on advanced mathematical models, MT protocols might be developed \textit{in silico}, leading to individualized treatment protocols with better clinical outcomes.

\newpage




\section{Tables and Figures}

\begin{table}[h!]
\centering
\caption{Characteristics (median and range) of the pediatric ALL population consisting of 116 (64 male and 52 female) patients. The body surface area was calculated using the Mosteller formula.}
\medskip
\begin{tabular}{l|r|r|l}
\rowGrey \textbf{Characteristic} & \textbf{Unit} & \textbf{Median} & \textbf{Range} \\ \toprule
Age                   & year  & 4.75   & 1.1--17.1 \\
Weight                & kg    & 22     & 10--90 \\
Height                & cm    & 112.45 & 80--182.7 \\
Body surface area$^*$ & m$^2$ & 0.82   & 0.47--1.98 \\
6MP daily dose        & mg    & 40     & 5--150 \\
MTX weekly dose       & mg    & 15     & 1.25--60 \\
ANC                   & \uGL  & 1.8    & 0.0--19.9 \\ \bottomrule
\end{tabular}
\label{tab:datasetMedian}
\end{table}


\begin{table}[h!]
\caption{Different dosing protocols for our \textit{in silico} simulation study.
Identical protocols for the administration of 6MP for ClinicalData and FittedModels with a median of the patient-individual average daily dosages of $43.15$$\pm$$10.5$~\umgms (minimum 15.8~\umgms, maximum 72.9~\umgms).}
\label{tab:Protocols}
\medskip
\begin{tabular}{r|l|r}
\rowGrey \textbf{Nr} & \textbf{Description} & \textbf{Short} \\ \toprule
1 & Collected clinical data & (ClinicalData) \\
2 & Fitted model based on patient's actual dosing & (FittedModels) \\
3 & Daily 6MP administration of 25~\umgms (50\% of AIEOP dosis) & (25~\umgms) \\
4 & Daily 6MP administration of 50~\umgms (AIEOP dosis) &  (50~\umgms) \\
5 & Daily 6MP administration of 75~\umgms (NOPHO/UK dosis) &  (75~\umgms) \\
6 & Daily 6MP administration of 100~\umgms (200\% of AIEOP dosis) & (100~\umgms) \\
\bottomrule
\end{tabular}
\end{table}


\begin{table}[h!] 
\centering
\caption{Results of parameter estimations for different models.
 Shown are model characteristics (data based on absolute neutrophil count [ANC] or white blood cell [WBC] count and PK models for 6MP and MTX), median and standard deviation in parentheses of individual root mean squared errors (RMSE), mean absolute errors (MAE), and final objective function values (FinalOBJ).}
\medskip
\begin{tabular}{l|rrrr}
\rowGrey
           & Model 1      & Model 2        & Model 3    & Model 3  \\ \toprule
Data       & ANC          & ANC            & ANC         & WBC          \\
PK 6MP     & Jayachandra  & Jayachandra    &   Hawwa     &  Hawwa    \\
PK MTX     & Panetta      & -              &    -        &    -        \\
		  \noalign{\smallskip}
MAE        &  1.068 (1.65) & 1.045 (1.92)    & 0.957 (4.31)  &  1.315 (2.92)               \\
RMSE        & 1.033 (0.49) & 1.022 (0.54)    & 0.978 (0.68)  &  1.147 (0.58)       \\
		  \noalign{\smallskip}
FinalOBJ   &  7002.68     & 7094.07        & 6550.48      &  9745.94$^*$   \\
\noalign{\smallskip}\hline \multicolumn{5}{l}{\begin{tiny}$^*$Objective value is not comparable to first three values due to different dataset\end{tiny}} 
\end{tabular} 
\label{tab:ModelComparison}
\end{table}



\clearpage
\begin{table}[h!]
	\centering
	\small
	   \caption{Model constants of the pharmacokinetic model of 6MP and its metabolite 6-TGN from \cite{Hawwa2008}, death rate constant of matured neutrophils, and initial conditions of the model \eqref{eq:FinalPKPDmodel}.}	\label{tab:Constants} 
\medskip
	\begin{tabular}{l l l l}
\rowGrey
		\textbf{Constant} & \textbf{Value} & \textbf{Unit} & \textbf{Description / Comment} \\ \toprule
		$ F $ & $ 0.22 $  & $   $ & Bioavailability factor\\
		$ k_a $  & $ 31.2 $ & 1/day & Absorption rate constant of 6MP \\
		$ k_{20} $ & $ 12.72 $ & 1/day  &  Elimination rate constant of 6MP\\
		$ \text{FM}_3 $  & $ 0.019 $&   &  Fractional metabolic transformation into 6TGN \\
		$ k_{me} $  & $ 9.9216 $ & 1/day & Metabolic transformation rate constant of 6MP \\
& & & into either 6TGN or 6MPN \\
		$ \text{CL}_{6tgn}(\text{BSA}) $  & $ 0.219 \ (\text{BSA})^{1.16} $ & L/day & Body surface area (BSA) dependent clearance of \\
& & & metabolite 6-TGN \\
              $k_{ma}$          & $2.3765$                  & 1/day & Death rate of matured neutrophils/leukocytes  \\
              $u(t_i)$          &                           &      mg            &  6MP amount at time point $t_i$  \\
              $\xOne(0)$ & 0  & mg & Same initial value for $\xTwo(0)$\\
              $\xThree(0)$ & 0  & \umgL & \\
              $\xFour(0)$ & $(\text{Base} \ \mkma)/\mktr$ & \uGL & Same initial value for $\xFive(0)=\xSix(0)=\xSeven(0)$ \\
              $\xEight$ & Base & \uGL & \\ \bottomrule
	\end{tabular}
\end{table}


\begin{table}[h!]                                                       
\centering
\caption{Results of parameter estimations of the final model using all (in-sample) or 70\% (out-of-sample) of the ANC values. 
Shown are parameter estimates of fixed effects, interindividual variability, and median errors of the parameter estimations. 
Relative standard errors are shown in parentheses.}
\medskip
\begin{tabular}{r|r@{\hskip 1.2mm}r|r@{\hskip 1.2mm}l}
\rowGrey
\textbf{Data}       & \multicolumn{2}{c|}{\textbf{In-sample}} & \multicolumn{2}{c}{\textbf{Out-of-sample}}   \\ \toprule 
\multicolumn{5}{c}{\textbf{Fixed effect parameters}}    \\\midrule
   Base             & 2.34   & (1.1)   &   2.06   & (0.1)  \\
   $\ktr$           & 0.148  & (0.4)   &   0.146  & (0.2)\\
   slope            & 0.242 & (0.2)   &   0.103 & (0.2)\\
   $\gamma$         & 0.769  & (0.1)   &   0.866  & (0.2) \\
\midrule 
\multicolumn{5}{c}{\textbf{Interindividual variability as coefficients of variation}}    \\  \midrule
   Base             & 23.1   & (18.9)  &   27.5   & (9.9) \\
    $\ktr$          & 16.5   & (25.6)  &   7.2    & (3.3)\\
    slope           & 44.9   & (5.4)  &   67.8   & (1.0)\\
    $\gamma$        & 10.7   & (0.5)  &   16.5   & (0.4) \\ 
\midrule 
 Proportional additive error   & 0.226 & (2.17) &  0.226 & (FIXED)  \\  
\midrule 
 \multicolumn{5}{c}{\textbf{Parameter estimation errors} }   \\ \midrule
Mean absolute error        & 0.957 & (4.31) & 1.466 & (490.30) \\
Root mean squared error    & 0.978 & (0.68) & 1.211 & (6.93) \\
\bottomrule
\end{tabular}
\label{tab:FinalParameterEstimates}
\end{table}


\begin{figure*}[h!]
\begin{center}
\resizebox{1.\textwidth}{!}{

}
\end{center}
\caption{Visual predictive check (VPC), derived by 1000 simulations with the final parameter estimates from the first column of Table~\ref{tab:FinalParameterEstimates}, for
circulating ANCs (\uGL) versus time (days).
Black dots are the measured ANCs. 
Black and blue lines show the median and 2.5th and 97.5th percentiles of measurements and model predictions, respectively. 
The shaded areas represent the 95\% confidence intervals around the 2.5th, 50th and 97.5th percentiles of the model predictions. 
Two ANC outliers (19.9 and 17.8) at time points 285.42 and 340.42 days are not shown.}
\label{fig:VPC}    
\end{figure*}



\clearpage

\section{Additional Requirements}


\section*{Conflict of Interest Statement}

The authors declare that the research was conducted in the absence of any commercial or financial relationships that could be construed as a potential conflict of interest.

\section*{Author Contributions}


FJ developed the population PK/PD models, performed the numerical computations and wrote the first draft of the manuscript.
JZ, TTTL, TR, MR, MM, and SS contributed to the model development, the study designs, and the interpretation of the results.
MSu and MSt provided clinical data.
All authors contributed to writing the final manuscript.

\section*{Funding}
This project has received funding from the European Research Council (ERC, grant agreement No 647573)
and from the European Regional Development Fund (grants SynMODEST and SynIsItFlutter)
under the European Union's Horizon 2020 research and innovation program.




\section*{Data Availability Statement}
The dataset is available upon reasonable request from the corresponding author.

\clearpage
\bibliographystyle{frontiersinHLTH&FPHY} 
\bibliography{popALL}





\end{document}